\documentclass[preprint,aps,tightenlines,showkeys,nofootinbib,superscriptaddress]{revtex4}
\usepackage{latexsym,amssymb,amstext,amsmath}
\usepackage[dvips]{graphicx}
\usepackage[dvips]{color}
\newcommand{\beq}{\begin{eqnarray}}
\newcommand{\eeq}{\end{eqnarray}}
\newcommand{\bea}{\begin{eqnarray*}}
\newcommand{\eea}{\end{eqnarray*}}
\newcommand{\eq}{eqnarray}

\newcommand{\al}{{\alpha}}
\newcommand{\be}{{\beta}}

\newcommand{\ga}{{\gamma}}

\newcommand{\ep}{{\epsilon}}

\newcommand{\de}{{\delta}}

\newcommand{\la}{{\lambda}}

\newcommand{\m}{{\mu}}
\newcommand{\n}{{\nu}}

\newcommand{\ka}{{\kappa}}

\newcommand{\Om}{{\Omega}}

\newcommand{\no}{{\nonumber}}
\newcommand{\f}{\frac}
\newcommand{\wt}{\widetilde}
\newcommand{\ra}{\rightarrow}

\newcommand{\vp}{\varphi}

\newlength{\wdth}

\begin{document}

\preprint{arXiv:2307.10532v2 [hep-th]}
\title{No Scalar-Haired Cauchy Horizon Theorem in Charged Gauss-Bonnet Black Holes}

\author{Deniz O. Devecio\u{g}lu \footnote{E-mail address: dodeve@gmail.com}
%}\affiliation{School of Physics, Huazhong University of Science and Technology,
%Wuhan, Hubei,  430074, China }
%\author{
and Mu-In Park \footnote{E-mail address: muinpark@gmail.com, Corresponding author}}
\affiliation{ Center for Quantum Spacetime, Sogang University,
Seoul, 121-742, Korea }
\date{\today}

\begin{abstract}
{Recently, a ``no inner (Cauchy) horizon theorem" for static black holes with
{\it non-trivial scalar hairs} has been proved in Einstein-Maxwell-scalar theories and also in Einstein-Maxwell-Horndeski
theories with {the} non-minimal coupling of
%the
{a charged {(complex)}} scalar field to
%the
Einstein tensor.
In this paper, we study an extension of the theorem to the static black holes in Einstein-Maxwell-Gauss-Bonnet-scalar theories, or simply,
charged Gauss-Bonnet (GB) black holes. We find that {\it no inner horizon with
charged scalar hairs} is allowed for the planar ($k=0$) black holes, as in
the case without GB term. On the other hand, for the non-planar ($k=\pm 1$) black holes, we find that {the}
haired inner horizon can not be excluded due to GB effect generally,
 %contrary to earlier works,
though we can not find a simple condition for {its}
%the
existence.
% of {the haired inner horizon}.
As {some} explicit examples of the theorem, we study
%some
numerical GB black hole solutions with {charged scalar} hairs and
%the
Cauchy horizons in asymptotically anti-de Sitter space{,} and find good
agreements with the theorem.
%As a byproduct,
{Additionally, in an Appendix, we %find
prove} a ``no-go theorem" for charged de Sitter
%GB
black holes {(with or without GB terms)} with charged scalar hairs
in arbitrary dimensions.
%We discuss
}
\end{abstract}

%\pacs{....}

\maketitle

\newpage
\section{Introduction}
{\label{section1}}

It is an important question in general relativity (GR) whether the predictability is lost due to {an} inner (Cauchy) horizon or not. Recently, {Cai {\it et al.} established \cite{Hart:2012,Cai:2020}} a ``no inner (Cauchy) horizon theorem" for both planar and spherical static black holes with {\it charged} {(complex)} scalar hairs in Einstein-Maxwell-scalar (EMS)
theories {that}
%which
suggests unstable Cauchy horizons in the presence of charged scalar hairs.
%, which is also known as {\it Q-hairs}.
A remarkable thing in the proof is
%the
{its} simplicity and quite generic results which do not depend on the details of the
scalar potentials {and} so
%that
it's applicability would be quite far-reaching.

As the first step towards the classification of all possible
extensions of the theorem, we extended the theorem to black holes in
Einstein-Maxwell-Horndeski (EMH) theories, with {the}
%a
non-minimal coupling of a {charged}
%complex
scalar field to
%the
Einstein tensor as well as the usual minimal gauge coupling \cite{Deve:2021}. Actually, since the Horndeski term is a four-derivative term via the Einstein coupling, it generalizes the recent set-up in \cite{Hart:2012,Cai:2020} into {\it higher-derivative} theories. There have been also several other generalizations of the theorem \cite{Yang:2021,An:2021,Cai:2021}.

In this {paper}, we {will} consider {an extension of} the theorem in another important higher-derivative gravity, Einstein-Maxwell-Gauss-Bonnet-scalar (EMGBS) theories {with {the} coupling of {a charged scalar hair} and the Gauss-Bonnet (GB) term \cite{Love:1971}}, whose equations of motion remain second order as in the Horndeski case. The
organization of {this} paper is as follows. In Sec. II, we consider the set-up for
%Einstein-Maxwell-Gauss-Bonnet-scalar (
EMGBS theories in arbitrary dimensions and obtain the equations of motion for a static and spherically symmetric ansatz. In Sec. III, we consider the radially conserved scaling charge and ``no scalar-haired Cauchy horizon theorem" for charged GB black holes. In Sec. IV, we consider the near horizon relations for the fields and the condition of scalar field at the horizons.
In Sec. V, we consider
%some
numerical black hole solutions with a Cauchy horizon in asymptotically
anti-de Sitter (AdS) space and find good agreements with the theorem.
In Sec. VI, we conclude with several remarks. {In Appendix {\bf D},
in addition, we prove a no-go theorem for charged de Sitter
%GB
black holes (with or without GB terms) with charged scalar hairs in arbitrary dimensions.}

\section{The action and equations of motion}
{\label{section2}}

As the set-up, we start by considering the $D$-dimensional
Einstein-Maxwell-Gauss-Bonnet-scalar (EMGBS)
action{, up to boundary terms}, with
%the
{a} $U(1)$ gauge {field} $A_{\mu}$ and {a} {\it charged} {(complex)} scalar field $\varphi$,
\begin{\eq}
I&=&\int d^D x \sqrt{-g} \left[ \f{1}{\kappa} R +{\cal L}_m \right], \label{action}\\
{\cal L}_m &=& -\f{Z(|\varphi|^2)}{4 \kappa} F_{\mu \nu} F^{\mu \nu}-\al (D_{\mu} \varphi)^* (D^{\mu} \varphi)-V(|\varphi|^2) \no \\
&&+\be (|\varphi|^2) \left(R_{\mu \nu \al \be} R^{\mu \nu \al \be} -{4 R_{ \al \be} R^{\al \be}}+R^2\right),
\end{\eq}
where $F_{\mu \nu} \equiv \nabla_{\mu} A_{\nu}-\nabla_{\nu} A_{\mu}$,
 %and
 $ D_{\mu} \equiv \nabla_{\mu}-i q A_{\mu}$ with {the} scalar field's charge $q$, and $\kappa \equiv 16 \pi G$. Here, we have introduced $Z$, $V$, and $\be$ as arbitrary functions of $|\varphi|^2$, as well as the usual minimal coupling $\al$. Note that the last term is {the} (quadratic) Gauss-Bonnet (GB) term \cite{Love:1971}, which is a topological, {\it i.e.}, total-derivative, term in $D=4$ for a constant coupling $\be$ so that it does not affect equations of motion and the local structure of spacetime.

Let us now consider the general {maximally}
%spherically
symmetric and static ansatz,
\begin{\eq}
ds^2&=&-h(r)dt^2+\f{1}{f(r)}dr^2+r^2 d \Omega_{D-2,k},
\label{Ansatz}\\
 \varphi&=&\varphi(r),~A=A_t(r)dt, \no
\end{\eq}
where $d \Omega_{D-2,k}$ is the metric of $(D-2)$-dimensional unit sphere with
%a
{the} spatial curvatures which are normalized as $k=1,0,-1$ for the spherical, planar, hyperbolic topologies, respectively. The equations of motion [Here, we consider $D=4$ case for simplicity but we will discuss about higher-dimensional cases later, which are straightforward  (for the details, see Appendix {\bf A})] are given by,
\begin{\eq}
E_{A_t}&\equiv& \f{1}{\ka} \left( Z(|\varphi|^2) \sqrt{\f{f}{h}}~ r^{2} A_t' \right)'
-\f{2 q^2 \al r^2 |\varphi|^2 A_t }{f} \sqrt{\f{f}{h}} =0,
\label{E_A}\\
E_{\vp}&\equiv& \left(\sqrt{\f{h}{f}}f \al r^2 \vp' \right)'
+\f{ q^2 \al r^2 A_t^2 \varphi  }{h}\sqrt{\f{h}{f}}   +r^{2} \left(\f{1}{2 \ka} \sqrt{\f{f}{h}} \dot{Z} {A_t'}^2-\sqrt{\f{h}{f}}~ \dot{V} \right) \vp+ \dot{\be} \vp {\cal G}=0, \label{E_varphi}\\
E_{h} &\equiv & \f{2}{\ka} \left(f-k+r f' \right)+r^2 V (|\varphi|^2) +f \al r^2 |\varphi'|^2 +\f{r^2}{2 h} \left( \f{1}{\ka} {Z}f  {A_t'}^2
+2 \al q^2 |\varphi|^2 A_t^2  \right) \no \\
&&+8\dot{\be} \left[(k-3f) f' |\varphi \vp'|+2 f(k-f) ( |\varphi'|^2+|\varphi \vp''|) \right]+32 \ddot{\be} f(k-f) |\varphi \vp'|^2=0,
\label{E_h}\\
E_f &\equiv& \f{2}{\ka} \left(f-k+r f' \left(\f{h'}{h} \right)\right)+r^2 V (|\varphi|^2) -f \al r^2 |\varphi'|^2 +\f{r^2}{2 h} \left( \f{1}{\ka} {Z}f  {A_t'}^2
-2 \al q^2 |\varphi|^2 A_t^2  \right) \no \\
&&+8\dot{\be} \f{f}{h}(k-3f) h' |\varphi \vp'|=0, \label{E_f}
\end{\eq}
where ${\cal G}$ is the GB density
%term
in $D=4$,
\begin{\eq}
{\cal G} \equiv \f{2}{\sqrt{f} h^{3/2}} \left[ (k-f) f h'^2
-h \left( (k-3f) f' h'+2(k-f) f h''\right)\right].
\end{\eq}
[The prime $(')$ and %the
dot $(\dot{~})$ denote the derivatives with respect to $r$ and $|\varphi|^2$, respectively]
Here, (\ref{E_A}), (\ref{E_varphi}) are the equations for the gauge field $A_t$
and {the} scalar field $\vp$, and (\ref{E_h}), (\ref{E_f}) are for
the metric {functions} $h,f$, respectively. As far as we know,
there is no exact solution for the {charged} GB black holes
{({\it i.e.}, with a gauge field $A_t$)} with {the}
%a black holecharge, , a
scalar field's charge $q$
and the field-dependent coupling $\be (|\varphi|^2)$. In this paper, we
will consider
%some
numerical solutions of the above non-linear {ODE}
%PDE
(\ref{E_A}) $\sim$ (\ref{E_f})
by {\it shooting method} and study the black hole{'s} interior space-time{,
as well as the exterior space-time}. But before that, in the next section we
first study the ``no scalar-haired Cauchy horizon theorem" to  see whether
it can be extended to the case with {the}
%a
GB coupling also.

\section{No scalar-haired Cauchy horizon theorem}
{\label{section3}}

In this section, we consider the ``no scalar-haired Cauchy horizon theorem" for charged GB black holes {with a charged {(complex)} scalar hair}. One of the key ingredients in the proof {of the theorem} is the existence of {the} {\it radially} conserved scaling charge,
\begin{\eq}
{\cal Q}&=& (D-2) r^{D-3} \sqrt{\f{f}{h}} \left[ \f{1}{\ka}  \left(r {h}'-2h
-r {Z} A_t {A_t}' \right)-\f{2 (D-3)}{r^2} f (r {h}'-2h) \left( (D-4) \be
+4 r |\vp \varphi'| \dot{\be} \right)  \right] \no \\
&&+k  \int ^r dr~ r^{D-4} \left\{ 2(D-2) (D-3) \sqrt{\f{h}{f}}\left[ \f{1}{\ka} -4\dot{\be} \left( |\vp (f' \vp'+2 f \vp '')| +2 f |\vp'|^2 \right)-16 \ddot{\be} f |\vp \vp'|^2  \right] \right. \no \\
&&+\f{(D-4)(D-3)(D-2)}{r \sqrt{f} h^{3/2}} \left[\be \left( r h (f'h'+2 f h'')-6 f' h^2-r f h'^2 \right) +4 \dot{\be}f h (r h'-6h) |\vp \vp'|\right] \no \\
&&\left.+\f{\sum^D_{n=5} (n-5) (n-4) (n-3)}{r^2 \sqrt{f h}}~ 8 \be \left[r f h'+2 (k-2f) h \right] \right\} ,\label{Q}
\end{\eq}
%{\strike{by recovering the full dimensional dependencies}}
where we have recovered the {full} dimensional dependencies
\footnote{We thank Yizhou Lu for an earlier collaboration in obtaining the
%general
full dimensional dependencies.}. From the equations of motion (\ref{E_A})
%,\ref{E_varphi},\ref{E_N},
-~(\ref{E_f}), one can prove that ${\cal Q}$ is radially conserved, {\it i.e., r-independent},
due to a remarkable
relation,
% {[Q: Will you check?]},
%${\cal Q}'=0$,
\begin{\eq}
{\cal Q}'&=&2 \sqrt{\f{f}{h}} \left[ r \left( \f{h}{f} {{\cal E}_f}'+\f{1}{2} \f{{h}'}{f} {\cal E}_f \right)
+\left( (D-2) \f{h}{f} +\f{r}{2} \f{{h}'}{f}\right) {\cal E}_h \right. \no \\
&&\left.+\f{r}{2} h |\vp'~ {\cal E}_\vp|+\f{1}{2} \left( (D-2) A_t + r A_t' \right) {\cal E}_{A_t} \right] =0, \label{Q_prime}
\end{\eq}
where {${\cal E}_f =E_f r^{D-4}/2,~{\cal E}_h =-E_h r^{D-4}/2,~ {\cal E}_\vp =2 E_\vp/\sqrt{f h}, ~ {\cal E}_{A_t} =- E_{A_t}\sqrt{h/f}$}
represent the (rescaled) equations of motion for the fields $f,h,\vp$, and $A_t$,
respectively.

For the {\it planar} topology, {\it i.e.}, $k=0$, the {\it local} charge ${\cal Q}$ is {the} Noether charge associated with a scaling symmetry
(for the details, see Appendix {\bf B})
%for a neutral scalar field with $q=0$ \cite{Feng:2015,
\cite{Gubs:2009}.
For {{\it non-planar} topologies}, {\it i.e.}, $k=+1$
(spherical topology),
%the
or $(k=-1)$ (hyperbolic topology), one can still {\it construct} a
radially conserved charge ${\cal Q}$ by including the {\it non-local}
parts with space integrals
%integrations
so that the non-conserved terms from the local parts are {exactly} canceled \cite{Cai:2020,Deve:2021,Yang:2021}.

Now, due to the $r$-independence of the scaling charge ${\cal Q}$, one can
consider the charges at the
%two
horizons, {in particular,}
%{\it i.e.,}
the outer event horizon $r_+$ and the inner Cauchy horizon $r_-$, if exists,
so that we have
%the relation
\begin{\eq}
&&\f{(D-2)}{\ka} \left[ r_+^{D-2} \left.\sqrt{\f{f}{h}}  \left({h}'-Z A_t A_t'  \right)\right|_{r_+}
-r_-^{D-2}  \left. \sqrt{\f{f}{h}}\left({h}'-Z A_t A_t'  \right)\right|_{r_-} \right] \no \\
&=&-k  \int ^{r_+}_{r_{-}} dr~ r^{D-4} \left\{ 2(D-2) (D-3) \sqrt{\f{h}{f}}\left[ \f{1}{\ka} -4\dot{\be} \left( |\vp (f' \vp'+2 f \vp '')| +2 f |\vp'|^2 \right)-16 \ddot{\be} f |\vp \vp'|^2  \right] \right. \no \\
&&+\f{(D-4)(D-3)(D-2)}{r\sqrt{f} h^{3/2}} \left[\be \left( r h (f'h'+2 f h'')-6 f' h^2-r f h'^2 \right) +4 \dot{\be}f h(r h'-6h) |\vp \vp'|\right] \no \\
&&\left.+\f{\sum^D_{n=5} (n-5) (n-4) (n-3)}{r^2 \sqrt{f h}}~ 8 \be  \left[r f h'+2 (k-2f) h \right] \right\}.
\label{Q_rel}
\end{\eq}

From the property
%of
$ {h}'\sqrt{{f}/{h}} |_{r_-} \leq 0$ at the assumed inner horizon $r_-$
%,
{and}
for non-extremal black holes with %a
{the} finite Hawking temperature or surface gravity
$\sim  {h}' \sqrt{{f}/{h}}|_{r_+} >0$ at the outer horizon
{$r_+$ with $f|_{r_{\pm}}\sim h|_{r_{\pm}}=0$}, the {purely metric-dependent} terms in the left-hand side of (\ref{Q_rel}) are {\it positive} {, {\it i.e.}, non-negative} (Fig. 1). Moreover, one can prove that $A_t$ needs to be zero at the horizons,
\begin{\eq}
A_t (r_+)=A_t (r_-)=0,
\label{BC_A}
\end{\eq}
from the regularity at the
horizons with ``charged" {(complex)} scalar fields \cite{Cai:2020,Deve:2021}
so that the gauge field terms in the left-hand side of (\ref{Q_rel}) vanish
at the horizons. This regularity condition is another key ingredient
%in
{for} the proof {of the theorem} (see Appendix {\bf C} for the proof {of (\ref{BC_A})}).

\begin{figure}
\includegraphics[width=10cm,keepaspectratio]{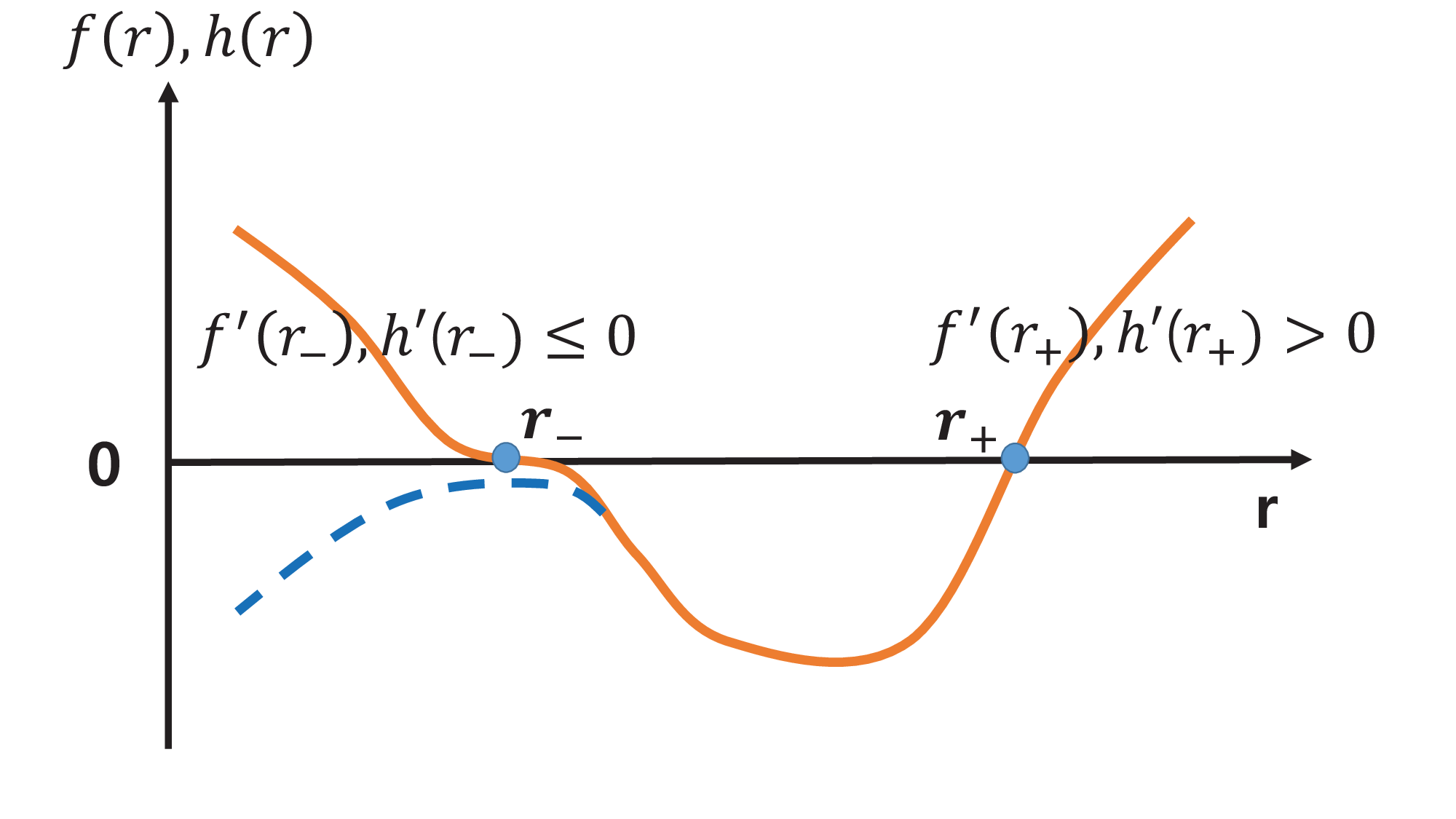}
\caption{Typical plots of $f(r)$ and $h(r)$ for a {\it non-extremal}
%{\it finite-temperature},
{charged} black hole with the inner (Cauchy) horizon $r_-$ and the outer (event)
horizon $r_+$. Due to {the} non-vanishing Hawking {temperature} at the Killing {horizon} $r_{\pm}$ with {$f|_{r_{\pm}}\sim h|_{r_{\pm}}=0$}, we have
$f'\sim {h}'>0$ at the outer horizon $r_+$, but
$f'\sim {h}' \leq 0$ at the inner horizon generally,
{\it if exists} (orange, solid line). In the absence of
{the}
%an
inner horizon,
 $f'$ and ${h}'$ can have arbitrary values but there is no inner point
 of $f\sim h=0$ (blue, dashed line).
%On the other hand, in the absence of outer horizon but now with an inner
%horizon, the situation is similar for an observer sitting inside the
%inner horizon which becomes now an event horizon  (green, dashed line).
}
\label{fig:N_f_AdS}
\end{figure}

Hence, the left-hand side of (\ref{Q_rel}) is always positive if the two horizons exist, which is consistent with the relation (\ref{Q_rel}) only if the right-hand side is also positive: This is a {\it necessary} condition (but not sufficient)
for the existence of the inner horizon with
%an
{the} outer horizon (or {the existence of} the outer
horizon with
%an
{the} inner horizon), from the radial conservation of the
scaling charge ${\cal Q}$. Otherwise, the assumption of existence of
%an
{the} inner (or outer) horizon with
%the
charged scalar hairs will not be true, which
%corresponds to
{results in} a {\it no-go theorem for the scalar-haired Cauchy horizon}. This is a powerful theorem since it does not depend on the details of scalar potential $V(|\vp|^2)$, contrary to the usual no-hair theorems.

{In addition, if there is another horizon, exterior to the outer
horizon $r_+$, {\it i.e.,} the cosmological horizon $r_{++}$ in asymptotically
de Sitter space, one can prove also a no-go theorem for charged de Sitter
%GB
black holes (with or without GB terms) with charged scalar hairs, since (\ref{BC_A}) can not be satisfied for both the outer horizon and the cosmological horizon (see Appendix {\bf D} for the proof).}

Then, for $k=0$, one can easily find that (\ref{Q_rel}) can not be satisfied since the right-hand side is zero which is in contradiction to the non-vanishing (positive) left-hand side. This means `` no smooth inner (Cauchy) horizon with the outer horizon" can be formed for black
{\it planes}.
% ($k=0$).

However, for $k\neq 0$,
%for other topologies, {\it i.e.,}
%$k=+1$ (black holes with spherical topology) and $k=-1$,
%(topological black holes with hyperbolic topology),
the right-hand side does not have a definite sign generally with the GB terms
($\be \neq 0$), and so there is no simple condition for the (non) existence of
%an
the inner (or outer) horizon associated with
{non-planar} topologies.
% parameter $k\neq 0$.
This implies that, for $k=+1$ (black holes with spherical topology) or $k=-1$ (topological black holes with hyperbolic topology), we may have a Cauchy horizon due to GB terms, depending on the solutions. For example, if we consider the $D=4$ case, for simplicity, where only the first integral in the right-hand side of (\ref{Q_rel}) remains, all the GB terms, except $|\vp (f' \vp'+2 f \vp '')|$, are positive for $\dot{\be}(|\vp|^2), \ddot{\be}(|\vp|^2)>0$ so that the no-inner-horizon theorem in EMS theories could apply. However, the excepted term can cause a problem in applying the theorem because of the sign-changing behaviors of $f'$ at the two horizons {with {$f|_{r_{\pm}}\sim h|_{r_{\pm}}=0$}}, {\it i.e.}, $f'|_{r_+}>0, f'|_{r_-}<0$, unless compensated by $\vp$, {\it i.e.}, $\vp \vp'|_{r_+}<0,~\vp \vp'|_{r_-}>0$, but $\vp \vp''|_{r_+}, \vp \vp''|_{r_-}>0$. In higher dimensions, we might need more complicated conditions.

This is in contrast to the EMS (Einstein-Maxwell-scalar)  theories
($\be=0$) \cite{Cai:2020} or EMH (Einstein-Maxwell-Horndeski)  theories
with a positive coupling $\ga>0$ to
%the
Einstein tensor \cite{Deve:2021}, where (\ref{Q_rel}) is inconsistent unless $k=-1$
%, {\it i.e.}, topological black holes with hyperbolic topology
so that ``no smooth inner (Cauchy) horizon with an outer horizon" can be
formed for $k=0$
% (black planes)
or $k=+1$.
 %(black holes with spherical topology).
However, even with GB terms,
we have the general criterion as follows, depending on the integral parts
of (\ref{Q_rel}): (a) if the right-hand side (RHS) of (\ref{Q_rel}) is zero
or negative, {the}
%an
inner (or outer) horizon can not be formed, (b) if RHS of (\ref{Q_rel}) is
positive, {the}
%an
inner (or outer) horizon can be formed but not necessarily always. We will see
{that} this general criterion {is satisfied by}
%in
the
%explicit examples of
numerical solutions
%of
in Sec. V.

%Before finishing this section, we note that one can write the RHS of
%(\ref{Q_rel}) in a more compact way as
%\begin{\eq}
%-k  \int ^{r_+}_{r_{-}} dr \sqrt{\f{h}{f}} {\cal F},
%\end{\eq}
%where ${\cal F}=2 [1/\ka -4\sqrt{f} (\sqrt{f} \be')']$. Then, the theorem says
%that no inner horizon can not be formed unless $k {\cal F}<0$, {\it i.e.}, $k=-1$
%with ${\cal F}>0$\footnote{} (as in EMS case) or $k=+1$ with ${\cal F}<0$.

%Here, it interesting to note that the theorem may also imply the
%{\it no outer horizon with an inner horizon} as described in
%Fig.1 (green, dashed line).
%This completes the proof of the theorem for the {\it charged} scalar
%field with $q \neq0$ and note that the proof
%is quite generic without assuming any specific form of the scalar potential.

\section{Near horizon relations}
%A regularity relations at the horizons
{\label{section4}}

In order to find numerical solutions for charged GB black holes, it is useful
to study
%about the
behaviors of metric functions, scalar and gauge fields near the horizons. To this end, we first solve {the equation for $f$} (\ref{E_f}), which is a quadratic polynomial of
$f$, and obtain
\begin{\eq}
f^{-1} \equiv e^B=
\f{-\mu(r) {\pm} \sqrt{\m(r)^2-4 \n(r)}}{2},
\label{eB}
\end{\eq}
where \footnote{
%In order to be compared with the literature about
For a real scalar field ($q=0$)
%and
with the GB coupling $\be(\vp)$, we may simply replace $2 \vp \dot{\be} (|\vp|^2)$ {in our results} by $\dot{\be} (\vp)$ \cite{Tori:1998,Anto:2017,Hunt:2020}.}
\begin{\eq}
\m(r)&=& \f{2 \left(4 k |\vp \vp'| \dot{\be}
%(|\vp|^2)
+{r}/{\ka} \right) A'
+\left({r^2} e^{-A} Z A_t'^2+{4} \right)/{2\ka}
-\al r^2 |\vp'|^2 }
{-\al q^2 r^2 e^{-A} |\vp|^2 A_t^2-2k/\ka+r^2 V},
%(|\vp|^2)},
\label{mu}\\
\n(r)&=& \f{-24 |\vp \vp'| \dot{\be}
%(|\vp|^2)
A'}{-\al q^2 r^2 e^{-A} |\vp|^2 A_t^2-2k/\ka+r^2 V
%(|\vp|^2)
}
\label{nu}
\end{\eq}
by introducing $e^A\equiv h(r)$ and $e^{B} \equiv f^{-1}(r)$ for convenience
\footnote{{There is no loss of generality, with $e^A$ and $e^{B}$, in analyzing
the black hole interior as well as the black hole exterior:
% $e^A, e^{B}>0$, solution.
One can still consider the black hole interior with $e^A, e^{B}<0$ by considering $e^A=-e^{\widetilde{A}}, e^{B}=-e^{\widetilde{B}}$ with $A=\widetilde{A}\pm i \pi, B=\widetilde{B}\pm i \pi$, and real functions ${\widetilde{A}}$ and ${\widetilde{B}}$.}}. Here, we note that the first term in the denominator $\sim e^{-A} |\vp|^2 A_t^2$ is finite and may not be neglected at the horizons generally{, with $e^{-A}\ra \pm \infty, A_t \ra 0, \vp =finite$}. Once the solutions for $A(r),\vp(r)$, and $A_t(r)$ are obtained,
the metric function $B(r)$ {can be {also} obtained by (\ref{eB}).} {The $\pm$ roots
depend on the topology parameter $k$ and the region {we}
%you
consider, {\it i.e.}, the black hole interior $r \in [r_-,r_+]$ or the black hole
exterior $r \in [r_+, \infty]$ or deep interior $r \in [0, r_-]$.}
%may be found easily.

By substituting $e^B$ from (\ref{eB}), the remaining
%independent
equations (\ref{E_A}), (\ref{E_varphi}), and (\ref{E_h}) with the three independent fields $A,\vp, A_t$ can be written as
\begin{\eq}
{A''=\f{P}{S}, ~\vp''=\f{U}{W}, ~A_t''=\f{R}{Y},\label{PUWRSY}}
\end{\eq}
while $B'$ can be written as
\begin{\eq}
B'(r)=-\f{\n' +e^B \m'}{2 e^{2B}+e^B \m}
\label{eB'}
\end{\eq}
{for both $\pm$ roots of $e^B$.} Here, {$P,S, U, W, R, Y$} are complicated functions of
{($r, e^B,e^{A}, A', \vp', A_t', \dot{\be}, \ddot{\be}$)}
%{(no $B'$ dependence)}
whose explicit expressions are not
%be
shown here (for the computational details, see Appendix {\bf E}). Since $A''$
diverges already at the horizons and $A_t''$ is finite for
%a
regular solutions with a finite $A_t'$ and the condition of $A_t=0$
(\ref{BC_A}) at the horizons, $\vp''$ gives the most information at the
horizons. For the solvability of the information, we consider the case
where the
%first
term $\sim e^{-A} |\vp|^2 A_t^2$ in the denominator of (\ref{mu}) and (\ref{nu}) can be neglected by considering {a rapidly-decaying gauge field at the horizons,} $A_t^2 \sim h^\de ~(\de>1)$, {whilst satisfying (\ref{BC_A}) also}. Then, one can find that
\begin{\eq}
\vp''=C(\vp, \vp', A_t',T)A' +{\cal O}(1)
\end{\eq}
and so, in order to have a regular scalar field with a {\it finite} $\vp''$ at the horizons, we need
\begin{\eq}
C(\vp, \vp', A_t',T)=0
\label{C=0}
\end{\eq}
at the horizons
{$r_H$, {\it i.e.,}} $A\ra \infty, A' \ra \pm \infty$
($+\infty$ for $r_+$, $-\infty$ for $r_{-}$). Here, $T \equiv h'=e^A A'$
corresponds to Hawking temperature (up to a finite factor) and given by
%{[Q: Is this only for + root of $e^B$?]}
\begin{\eq}
T
%(r)
=\left. \f{\ka r^2 Z A_t'^2 \left[2 \ka r^2 Z \left(\al q^2 r^2 e^{-A} |\vp|^2A_t^2+2 \ka {k}-r^2 V\right)-\widetilde{Q}^2 (r) \right]}{4 \widetilde{Q}^2 (r)(4k |\vp \vp'| \dot{\be}+\ka r)} \right|_{r_H},
\label{T}
\end{\eq}
by keeping $e^{-A} |\vp|^2A_t^2$ term for the sake of completeness, which
will be neglected in the
%following
{next section for numerical solutions}, for simplicity.
(See Appendix {\bf E} for some relevant computations and the explicit form
of $C(\vp, \vp', A_t',T)$) $\widetilde{Q}(r)$ is {the}
%a
black hole charge
function
%and $\widetilde{Q}(r_H)$ is its integration constant
{by integrating} the gauge field equation (\ref{E_A}),
\begin{\eq}
&&\f{Z}{\ka} e^{-(A+B)/2} r^2 A_t'={\widetilde{Q}(r)}, \label{Q_tilde}\\
&&{\widetilde{Q}(r)}=2 \al q^2 \int^r_{r_H} dr r^2 e^{(B-A)/2} |\vp|^2 A_t + \widetilde{Q}(r_H)
\end{\eq}
so that {we obtain, from (\ref{Q_tilde}),}
\begin{\eq}
e^B=\f{\ka^2 Z^2 r_H^4}{\widetilde{Q}^2(r_H)}{A_t'}^2 e^{-A}
\label{eB_Q}
\end{\eq}
at the black hole horizons $r_H$: By comparing (\ref{eB_Q}) and the near horizon expansion of (\ref{eB}) with $T=e^A A'$, one can find the temperature formula (\ref{T}).

Then, solving the condition (\ref{C=0}), we obtain the condition of $\vp'|_{r_H}$
at the horizons from a finiteness of $\vp''|_{r_H}$. We note that the condition
of $\vp'|_{r_H}$ is important in obtaining
%the
numerical solutions in the next section because it gives the proper initial condition at the horizons. As we have noted above, we will consider the rapidly vanishing gauge field $A_t$ at the horizons so that $e^{-A} |\vp|^2 A_t^2$ term in (\ref{mu}), (\ref{nu}), and (\ref{T}) are neglected for the solvability of (\ref{C=0}).

\section{Numerical solutions}
{\label{section5}}

In Sec. \ref{section3}, we have proved that, for the {planar} topology $(k=0)$
{of charged GB black holes}, {the} Cauchy horizon with charged scalar fields can
not be formed. However, for
%other
{non-planar} topologies, {\it i.e.}, spherical ($k=1$) or hyperbolic ($k=-1$) black holes, there is no simple condition for the existence of the haired Cauchy horizons due to the GB term, except a general criterion on the integral parts of the
scaling charge. In this section, we consider
%show some
numerical solutions with the scalar-haired Cauchy horizon for the hyperbolic
case as
%an
some explicit examples of our theorem.

\begin{figure}
\includegraphics[width=8cm,keepaspectratio]{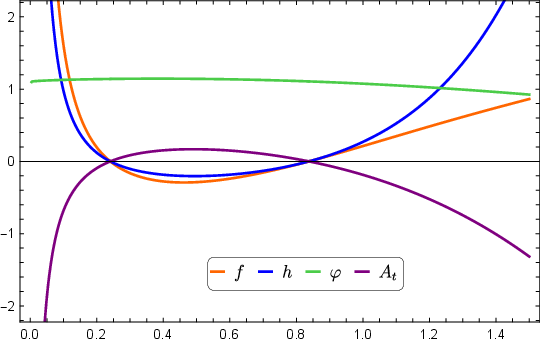}~
\includegraphics[width=8cm,keepaspectratio]{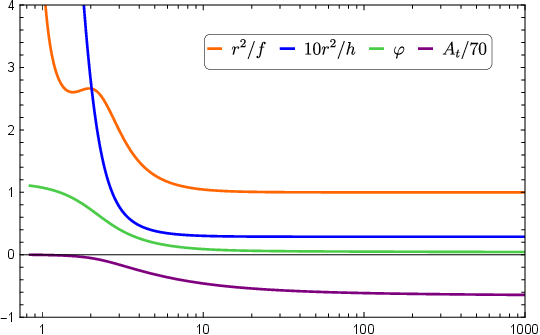}
\caption{Numerical solutions of the $D=4$ hyperbolic {charged} GB black hole
($k=-1$) with the model (\ref{model: Einstein_limit}) and initial
conditions (\ref{BC: Einstein_limit}) in asymptotically {\it AdS}
space. In the left panel, we plot $f$ (orange), $h$ (blue), $\vp$ (green),
$A_t$ (purple), and they show
%an
the inner
Cauchy horizon at $r_-\approx 0.240558232$ as well as
%an
the outer event horizon at $r_+ \approx 0.837565833$,
where all the solutions are smooth. We find $A_t=0$ at the {two} horizons,
{in consistent with}
%as
%described
%{given} by
the condition (\ref{BC_A}). In the right panel, we plot $r^2 /f,10 r^2/h,
{\vp, A_t/70}$, and the
solutions reveal the asymptotically {\it AdS-like} behaviors
$f \sim h \sim r^{2}$
%(after a constant rescaling of the time coordinate $t$)
with {$\vp \sim 0$} and a finite $A_t$.
%, as $r$ approaches to the asymptotic boundary.
}
\label{fig:Einstein_limit_Sol}
\end{figure}

\begin{figure}
\includegraphics[width=8cm,keepaspectratio]{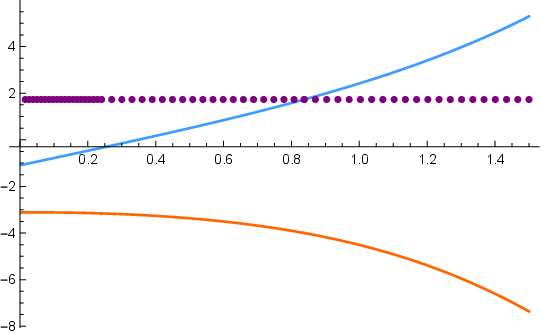}
\caption{Plots of the scaling charge ${\cal Q}$ (purple) together with its local (blue) and integrand {(orange)} parts  for the numerical solutions in Fig. \ref{fig:Einstein_limit_Sol}. The local and integrand parts are smooth at the horizons. We can see also the radial conservation, {\it i.e.}, constancy, of the charge ${\cal Q}$, across the horizons.
}
\label{fig:Einstein_limit_Q}
\end{figure}

In Fig. \ref{fig:Einstein_limit_Sol}, we first present numerical solutions of the four-dimensional {($D=4$)} hyperbolic {charged} GB black hole in EMGBS gravity for the choice of the model
\begin{\eq}
\be(|\vp|^2)=\la |\vp|^2,~V(|\vp|^2) =-6 +m^2 |\vp|^2, ~ Z(|\vp|^2)=1
\label{model: Einstein_limit}
\end{\eq}
with $\la =10^{-10}$, $m^2=-0.18388$ \footnote{We choose the same
{tachyonic mass
%, {\it i.e.},
($m^2<0$) of the}
%The chosen
scalar field
%'s mass
as in \cite{Cai:2020}.
 %which is .
 %But,
 Since it satisfies the Breitenlohner-Freedman (BF) bound $m^2 \geq -9/4$, it would be stable under small fluctuations in the global {\it AdS} background \cite{Brei:1982}.}, and $q=1.5$. We choose the outer horizon $r_+ \approx 0.837565833$ and the initial conditions at $r_+$ as follows (we have set $\ka \equiv \al \equiv 1)$:
\begin{\eq}
&&h(r_+)=-10^{-10},~h'(r_+)=1.23278,~A_t(r_+)=10^{-10},~A_t'(r_+)\approx -0.927989424, \no \\
~ &&\vp(r_+)=1.106834110,~ \vp'(r_+)=-0.165094, ~\widetilde{Q}(r_+)=0.650999915.
\label{BC: Einstein_limit}
\end{\eq}

Here, $\vp'(r_+)$ and $A_t'(r_+)$ are determined by (\ref{C=0}) and (\ref{T})
by $T=h'(r_+)$, respectively. By rewriting the second-order equations of
motion (\ref{E_A}), (\ref{E_varphi}), (\ref{E_h}) into the first-order forms
via new variables
$H(r)\equiv h'(r),~ \Phi(r)\equiv\vp'(r),~ E(r) \equiv A_t' (r)$
(after replacing $f(r)$ and $f'(r)$ by (\ref{eB}) and (\ref{eB'}), respectively), we
numerically solve the six differential equations (with $k=-1$) {by {\it shooting} method,} for the variables $(h, H, \vp, \Phi, A_t, E)$ from $r=r_+ +\ep$ to $r=10^4$ for the black hole exterior solution (we set $\ep=10^{-9}$), and from
$r=r_+ -\ep$ to $r=10^{-8}$ for the black hole interior solution. We use
\texttt{NDSolve} of {MATHEMATICA}
% athematica}
with
\texttt{PrecisionGoal $\ra 26$} and \texttt{WorkingPrecision $\ra 27$}.

The {code for the} black hole interior solution
%in the Fig. \ref{fig:Einstein_limit_Sol} (left panel)
{solves the differential equations up to}
%shows
%an
{the} inner Cauchy horizon at $r_-\approx 0.240558232$ but
{does not solve beyond}
%do not show the deep interior solution behind
the inner horizon, due to increasing numerical errors at $r_-$. In order to obtain the deep interior solution {beyond the inner horizon}, we {need to consider another set up of the code with now} the initial conditions at $r_-$ as follows, which can be determined from the obtained numerical interior solutions:
\begin{\eq}
&&h(r_-)=10^{-10},~h'(r_-)=-2.33135,~A_t(r_-)=-10^{-10},~A_t'(r_-)\approx 1.868563472, \no \\
~ &&\vp(r_-)=1.140835529,~ \vp'(r_-)=-0.0560344, ~\widetilde{Q}(r_-)=-0.137024.
\end{\eq}

Here, in order to obtain the proper initial conditions of $h'(r_-)$ and  $\vp'(r_-)$, we first determine {the black hole charge} $\widetilde{Q}$ from $h'(r_-)=-2.328970925, \vp'(r_-)=-537884.8323$ which can be read {\it naively} from the interior solution, and use (\ref{C=0}) and (\ref{T}). It is important to note that $\wt{Q}$ is a quite stable parameter in the numerical solutions and its first determination gives the reasonable initial conditions of $\vp'(r_-)$ which matches well with the interior solution, even with the high values of {\it{naively-read}} $\vp'(r_-)$ due to numerical errors at $r_-$.

By combining all the solutions in three different regions, we obtain the whole region
solution as in Fig. \ref{fig:Einstein_limit_Sol} (left panel), which shows the smooth matchings at the inner and outer horizons. Here, note that we need to use the `$+$' root of $f^{-1}=e^B$ in (\ref{eB}) for the black hole interior solution, and the `$-$' root of for the deep interior and black hole exterior solutions, respectively. The exterior solutions in Fig. \ref{fig:Einstein_limit_Sol} (right panel) show the
asymptotically {\it AdS-like} behaviors $f \sim h \sim r^{2}$ with {$\vp \sim 0$}
and finite $A_t$.
%, after a constant rescaling of the time coordinate $t$.
Fig. \ref{fig:Einstein_limit_Q} shows
%the agreements of
the radially conserved ({\it i.e.}, constancy) of scaling charge, and its local and integrand parts {which show smooth matchings} at the horizons and so supports {well} our numerical solutions.

\begin{figure}
\includegraphics[width=8cm,keepaspectratio]{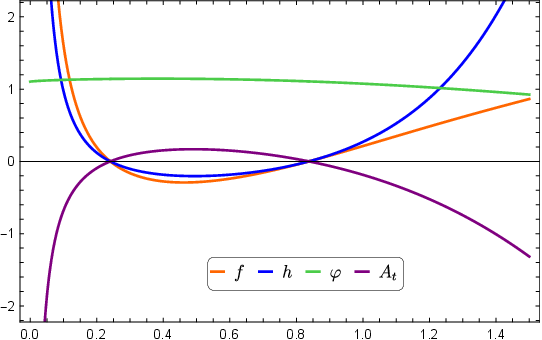}~
\includegraphics[width=8cm,keepaspectratio]{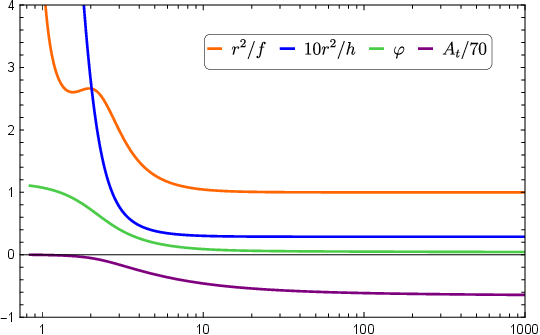}
\caption{Numerical solutions of a $D=4$ hyperbolic {charged} black hole
%($k=-1$)
in EMS gravity with
the same model (\ref{model: Einstein_limit}) and initial conditions (\ref{BC: Einstein_limit}) in asymptotically {\it AdS} space. This agrees {well} with Fig. \ref{fig:Einstein_limit_Sol} and also the earlier result in \cite{Cai:2020}, but {\it without} the {\it complexified} integration to avoid the coordinate singularity problem at the inner horizon.
}
\label{fig:Einstein_Sol}
\end{figure}

\begin{figure}
\includegraphics[width=8cm,keepaspectratio]{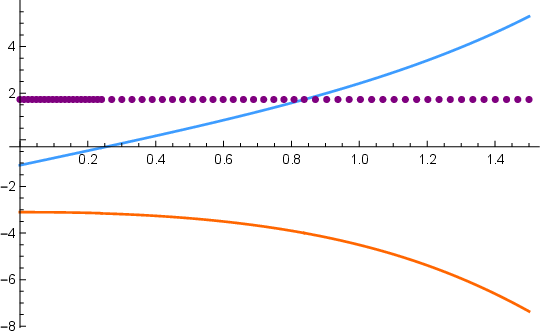}
\caption{Plots of the scaling charge ${\cal Q}$ (dots) and its local (solid line), integrand (dashed line) parts for the numerical solutions in Fig. \ref{fig:Einstein_Sol}.
}
\label{fig:Einstein_Q}
\end{figure}

In Fig. \ref{fig:Einstein_Sol} and \ref{fig:Einstein_Q}, for a comparison, we present numerical solutions of the corresponding hyperbolic {charged} black hole in the EMS case ($\be=0$) for the same choice of the model (\ref{model: Einstein_limit}) and initial conditions (\ref{BC: Einstein_limit}), as in \cite{Cai:2020}.
Fig. \ref{fig:Einstein_Sol} shows quite good agreements with
Fig. \ref{fig:Einstein_limit_Sol} which can be considered as EMS limit
(${\be} \ra 0$) so that it can be a consistency {check}
%test
of our numerical method. Actually, we reproduce the numerical solutions
in \cite{Cai:2020} (Fig. 3) but without using the {\it complexified} integration
to avoid the coordinate singularity at the inner horizon which can not be
applied to our case with GB term, due to different roots of
$f^{-1}=e^B$ (\ref{eB}) in different regions: There is only one root of $e^B$
in (\ref{eB}) without GB term
%({\it i.e.}
($\be=0$) and one has the same field equations for the whole region.
{One difference is the last point of the deep interior solution where
the numerical computation stops, $r\approx 0.0000262$, which is
smaller than $r\approx 0.0038097$ for Fig. \ref{fig:Einstein_limit_Sol} {and}
%, which
might indicate a possible GB effect near $r=0$.}

\begin{figure}
\includegraphics[width=8cm,keepaspectratio]{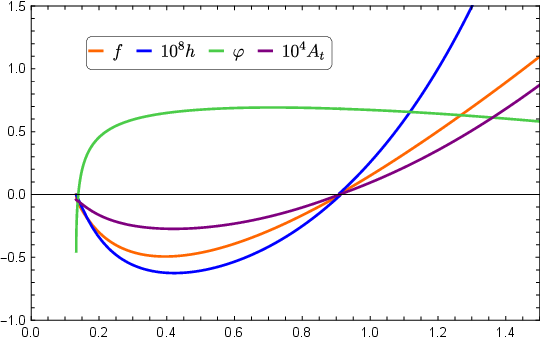}
\includegraphics[width=8cm,keepaspectratio]{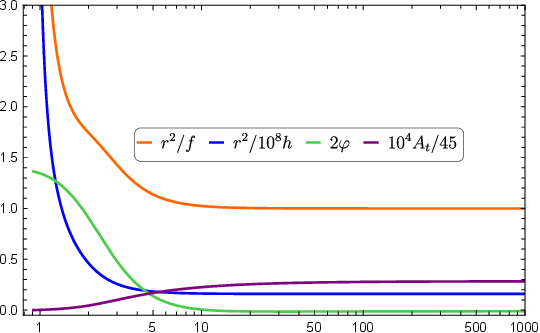}
\caption{Numerical solutions of a $D=4$ hyperbolic {charged} GB black hole
%($k=-1$)
with
the model (\ref{GBmodel:a})
%, (\ref{GBmodel:b}),
and initial condition (\ref{BC:GB_a}).
%, (\ref{BC:GB_b}).
In the left panel, we plot $f$ (orange), $10^8 h$ (blue), $\vp$ (green),
$10^4 A_t$ (purple), and no solution is found for the deep interior region due to
high numerical errors at the inner horizon $r_-\approx 0.132916515$. In the right
panel, we plot  $r^2/f,r^2/(10^8 h),{2 \vp}, 10^4 A_t/45$, and {they}
%it
show the {\it AdS-like} behaviors in $f$, $h$, $\vp\sim 0$ and finite $A_t$ as in Fig. \ref{fig:Einstein_limit_Sol}.
}
\label{fig:GB_Sol}
\end{figure}

\begin{figure}
\includegraphics[width=8cm,keepaspectratio]{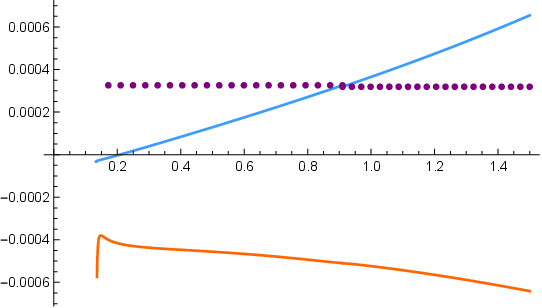}
\caption{Plots of the scaling charge ${\cal Q}$ (purple) and its local (blue), integrand (orange) parts for the numerical solution in Fig. \ref{fig:GB_Sol}.
}
\label{fig:GB_Q}
\end{figure}

In Fig. \ref{fig:GB_Sol}, we present another numerical solution of the $D=4$ hyperbolic {charged} GB black hole in EMGBS gravity for the model with
\begin{\eq}
\la =10^{-3},~m^2=-0.18,~q=2.5.
\label{GBmodel:a}
\end{\eq}
%for (a) and (b),
%\begin{\eq}
%\la =10^{-4},~m^2=-0.18,~q=1.5.
%\label{GBmodel:b}
%\end{\eq}
%for (c) and (d).
We choose initial conditions
\begin{\eq}
&&r_+=0.91, h(r_+)=-10^{-10},~h'(r_+)=2.48104 \times10^{-8},~A_t(r_+)=10^{-10},~A_t'(r_+)=10^{-4}, \no \\
~ &&\vp(r_+)=0.682,~ \vp'(r_+)=-0.0918761, ~\widetilde{Q}(r_+)=0.65.
\label{BC:GB_a}
\end{\eq}
%for (a) and (b),
%\begin{\eq}
%&&r_+=0.92, h(r_+)=-10^{-10},~h'(r_+)=2.77748 \times10^{-8},~A_t(r_+)=10^{-10},~A_t'(r_+)=10^{-4}, \no \\
%~ &&\vp(r_+)=1.10306,~ \vp'(r_+)=-0.123005, ~\widetilde{Q}(r_+)=0.65
%\label{BC:GB_b}
%\end{\eq}
%for (c) and (d).

Using the same method as in Fig. \ref{fig:Einstein_limit_Sol}, we obtain the
numerical solutions for the black hole interior and exterior
% solutions
as {shown} in Fig. \ref{fig:GB_Sol}. But, in this case
%ese cases,
no deep interior solution is
%s ar
found due to higher numerical errors at the inner horizon $r_-\approx 0.132916515$. For the exterior solution, we can see the similar {\it AdS-like} behaviors in $f(r)$, $h(r)$, $\vp \sim 0$ and finite $A_t$ as in Fig. \ref{fig:Einstein_limit_Sol}. Fig. \ref{fig:GB_Q} shows the radial conservation of ${\cal Q}$ as well as its local and integrand parts {with smooth matchings at the horizons} and {so} supporting our numerical solutions.

In all the numerical solutions, we find that (a) the vanishing gauge field condition (\ref{BC_A}), $A_t (r_H)=0$ and (b) the general criterion for the existence of an inner Cauchy horizon, {\it i.e.}, negativeness of RHS in (\ref{Q_rel}) (or integrand of ${\cal Q}$ in Figs. \ref{fig:Einstein_limit_Sol}, \ref{fig:Einstein_Q}, and \ref{fig:GB_Q}) are satisfied. {On the other hand, at the asymptotic boundary ($r=10^4$ in our analysis), we find that $r^2/f \approx 1, \vp \approx 0$, $r^2/h$ and $A_t$ approach to finite values with vanishing derivatives. This indicates the {\it AdS-like} behaviors of $f, h  \sim r^2, A_t \sim 1/r, \vp \sim 1/r^a $ with
$a>0$ \footnote{From the leading terms in large r {expansion} of the scalar field
equation (\ref{E_varphi}), we obtain
$a=\left(3\pm \sqrt{-96 {\la}+9+4 m^2}\right)/2$ for the model
(\ref{model: Einstein_limit}) ($\ka,\alpha \equiv 1$), which generalizes
the BF's result to GB gravity \cite{Brei:1982}. Our numerical solutions
correspond to $a_-\approx 0.06, a_+\approx 2.94$ (Fig. 2, 4) and
$a_-\approx 0.07, a_+\approx 2.93$ (Fig. 6) with the leading power
$a=a_-$. {The relatively larger values of the scalar fields ($0.0377$ for
Fig. 2 (or 3), $0.0048$ for Fig. 6) at our largest but finite $r$ boundary
will be due to their slowly-varying behaviors with the leading
%small
power of $a_-$.} However, there are some unusual complications in obtaining the full large r expansion, due to double roots of $a$, {and its} consistent asymptotic solution is still unclear. }.}

\section{Concluding Remarks}
{\label{section6}}

In conclusion, we have studied an extension of the ``no scalar-haired inner (Cauchy) horizon theorem" to the electrically charged Gauss-Bonnet (GB) black holes in Einstein-Maxwell-Gauss-Bonnet-scalar (EMGBS) theories {with {charged (complex)} scalar fields}. We have found that the {condition of} vanishing gauge field $A_t=0$ at the horizons from the regularity of equations of {motion} is unchanged even with GB coupling. We have computed the radially conserved scaling charge ${\cal Q}$ for GB coupling $\be(|\vp|^2)$ in arbitrary dimensions $D$.

From the constancy of scaling charge ${\cal Q}$ at the outer and inner horizons, we have obtained the {\it no scalar-haired Cauchy horizon theorem} for $k=0$, {\it i.e.}, ``no scalar-haired inner (Cauchy) horizon for planar ($k=0$) black holes". For $k=1$ (spherical) or $k=-1$ (hyperbolic) black hole, a simple condition for the no scalar-haired theorem is not available due to some terms in the integral part of ${\cal Q}$ which do not have definite signs in the black hole interior region, but one can still consider a general criterion for applying the theorem, depending on the resulting sign of the integral part. This means that a Cauchy horizon with
{\it non-trivial}, charged scalar hairs {{\it might}} exist {even}
for $k=1$,
%or {might} not exist for $k=-1$,
contrary to Einstein-Maxwell-scalar (EMS) or Einstein-Maxwell-Horndeski (EMH) theories, {as a genuine GB effect.}

We have obtained
%some
numerical solutions with the inner horizon for $k=-1$ and in the asymptotically anti-de Sitter space. For a very small GB coupling parameter $\la=10^{-10}$, we have obtained a solution covering the whole region of $r \in [0, \infty]$, which agrees with EMS solution of \cite{Cai:2020}, but without the ``complexification" method to obtain the deep interior solution of $r \in [0, r_-]$. On the other hand, for $\la= 10^{-3}$, we have obtained solutions for the inner and exterior regions, {\it i.e.},
$r \in [r_-, \infty]$ from the
%boundary data
{initial conditions} at the outer horizon $r_+$, but {no solution is found}
%have not
%succeeded {obtained}
for the deep interior region of $r \in [0, r_-]$ due to increased errors in
finding
%the boundary data
{initial conditions} at the inner horizon. We have checked the {radial conservation, {\it i.e.},} constancy of the scaling charge ${\cal Q}$ and the general criterion of the scalar-haired Cauchy horizon in the obtained numerical solutions.

{However, we were not able to find numerical solutions with the inner horizon for $k=1$, which might exist due to a GB effect. From the no-go theorem for $k=1$ in EMS theories \cite{Hart:2012,Cai:2020}, we  expect that the solution may exist in the {\it non-GR} branch, where the GB coupling is not small. On the other hand, in Appendix {\bf D},
we have shown a no-go theorem for charged de Sitter
black holes (with or without GB terms) with charged scalar hairs. }

%Some open problems are as follows: (1)
It would be a challenging problem to obtain {the whole range solutions including} the deep interior solutions for a sizable coupling $\la \sim {\cal O} (1)$ {also,} by controlling the numerical errors {at the inner horizons} more efficiently. There is no known exact {\it charged} GB black hole solution with {\it complex} scalar hairs as far as we know. It would be interesting to see whether some exact haired solutions can be found by considering some special potential and to study the GB effect on the interior and deep interior spaces.

%{\it Note added}: After finishing this paper, a related paper \ci{Bell:2019} appeared which is
%overlapping with ours

\section*{Acknowledgments}
We would like to thank
%Rong-Gen Cai,
Li Li
%, Tsutomu Kobayashi, and Kyung Kiu Kim
for helpful discussions and Yizhou Lu for an earlier collaboration. DOD was supported by Basic Science Research Program through the National
Research Foundation of Korea (NRF) funded by the Ministry of
Education, Science and Technology {(
%2016R1A2B401304,
2020R1A2C1010372)}.
%the National Natural Science Foundation of China under Grant No. 11875136
%and the Major Program of the National Natural Science Foundation of China
%under Grant No. 11690021.
MIP was supported by Basic Science Research Program through the National
Research Foundation of Korea (NRF) funded by the Ministry of
Education, Science and Technology {(
%2016R1A2B401304,
2020R1A2C1010372, 2020R1A6A1A03047877)}.

\appendix{\label{app1}}
\begin{section}
{Full equations of motion in arbitrary dimensions}
\end{section}
In this Appendix, we present the reduced action {for the {general static}
%spherical symmetric
ansatz (\ref{Ansatz})} and full equations of motion in arbitrary dimensions $D$,
{generalizing}
%for
the $D=4$ case (\ref{E_A})-(\ref{E_f}) in the text. First, the reduced action reads, {up to boundary terms},
\begin{\eq}
I&=&\int d^D x \sqrt{-g}\left(\frac{1}{\kappa}R+\mathcal{L}_{m}\right)\no\\
&=&\Om_{D-2,k} \int dt dr \Big[ \frac{r^{D-3}}{\sqrt{f h}} \Big\{
\frac{(D-2) h}{\ka r} \left((D-3) (k-f)-r f'\right)+\frac{r f  Z(|\vp|^2)}{2 \kappa}  A_{t}'^2 \no\\
&&+ \alpha    q^2 r |\vp|^2 A_{t}^2 -\alpha  r f h |\vp'|^2-r h V(|\vp|^2)
\Big\}
%\no\\ &&
+ \beta(|\vp|^2)~ \mathcal{G}_{1}+ |\vp \vp'| \dot\beta(|\vp|^2)~ \mathcal{G}_{2} \Big],
\end{\eq}
where
\begin{\eq}
\mathcal{G}_{1}&\equiv &  (D-4)(D-3)(D-2)  \sqrt{\frac{ h}{f}} r^{D-6} (f-k)  \left[(D-5) (f-k)+2 rf'\right],\\
\mathcal{G}_{2}&\equiv & 4 (D-3) (D-2)  \sqrt{\frac{f}{h}} r^{D-4}  (k-f) h'
\end{\eq}
{are the contributions from the GB term and $\Om_{D-2,k}$ is the volume of
$(D-2)$-dimensional hypersurface for a curvature parameter $k$.}
Varying the above action %with will
{leads} to the following equations for $A_{t},\vp,h$, and $f$
\begin{\eq}
E_{A_t}&\equiv& \f{1}{\ka} \left( Z(|\varphi|^2) \sqrt{\f{f}{h}}~ r^{D-2} A_t' \right)' -\f{2 q^2 \al r^{D-2} |\varphi|^2 A_t }{f} \sqrt{\f{f}{h}}  =0,
\label{E_A_Full}\\
E_{\vp}&\equiv& \left(\al r^{D-2} \vp'  f \sqrt{\f{h}{f}} \right)'+  \f{ q^2 \al r^{D-2} A_t^2 \varphi  }{h} \sqrt{\f{h}{f}}  \no \\
&&+r^{D-2} \left(\f{1}{2 \ka} \sqrt{\f{f}{h}} \dot{Z}(|\vp|^{2}) {A_t'}^2-\sqrt{\f{h}{f}}~ \dot{V}(|\vp|^{2}) \right) \vp+ \vp\dot\beta(|\vp|^2) \left(\mathcal{G}_{1}-\frac{1}{2} {\mathcal{G}_{2}'}\right) =0, \label{E_varphi_Full}\\
E_{h} &\equiv & \f{(D-2)(D-3)}{\ka} \left(f-k+\f{r f'}{D-3}\right)+r^2 V (|\varphi|^2) +\al f r^2|\varphi'|^2 \no \\
&&+\f{r^2}{2 h}\left( \f{1}{\ka} {Z(|\vp|^{2})}f {A_t'}^2
+2 \al q^2 |\varphi|^2 A_t^2   \right)\no\\
&&+\frac{2 \sqrt{fh}}{r^{D-4}} \left( \frac{|\vp\vp'| \dot{\beta}(|\vp|^2)}{h'(r)}\mathcal{G}_{2} \right)^{\prime}-\frac{1}{r^{D-4}} \sqrt{\frac{f}{h}} \left(\beta(|\vp|^2)\mathcal{G}_{1} -|\vp \vp'| \dot{\beta}(|\vp|^2)\mathcal{G}_{2} \right)=0, \no \\
\label{E_N_Full}\\
E_f &\equiv& \f{(D-2)(D-3)}{\ka} \left( f-k+\f{r f}{D-3} \left( \f{{h}'}{h}\right) \right)+r^2 V (|\varphi|^2)-\al f  r^2 |\varphi'|^2 \no \\
&&+\f{r^2}{2 h}\left( \f{1}{\ka} {Z(|\vp|^{2})}f {A_t'}^2
-2\al q^2  |\varphi|^2 A_t^2  \right) \no \\
&&+\left( |\vp \vp'| \dot{\beta}(|\vp|^2) {\cal H}_1+ \beta(|\vp|^2) {\cal H}_2 \right) \mathcal{G}_{2}=0,
\label{E_f_Full}
\end{\eq}
%with the definitions for $H$ and $P$
where
\begin{\eq}
{\cal H}_1&\equiv& \sqrt{\frac{f}{h}} \frac{ \left[2 (D-4) h (k-f)+r (k-3 f) h'\right]}{r^{D-3} (k-f) h'},\\
{\cal H}_2&\equiv& \frac{(D-4) }{4 r^{D-2} h'}\left[\sqrt{\frac{h}{f}}(D-5)  \left(f-k\right)
+\sqrt{\frac{f}{h}}2 r h'\right].
\end{\eq}
{We note that, for $D=4$, there is only the ${\cal G}_2$ term with a non-vanishing coefficient $\dot{\beta} (|\varphi|^2)$, whereas for higher dimensions $D>4$, there are both ${\cal G}_1$ and ${\cal G}_2$ terms with an arbitrary $\beta (|\varphi|^2)$. On the other hand, there is no GB contribution for the lower dimensions, {\it i.e.} $D=3, 2$. }

\section{Computational details of scaling charge formulas (\ref{Q}) and (\ref{Q_prime})}
{In this Appendix, we present computational details of the scaling charge formula (\ref{Q}) and its radial conservation equation (\ref{Q_prime}).}
Let us first start with the action \eqref{action} and, for the sake of computational simplicity, consider the planar metric with the following coordinate choice,
\begin{align}
ds^{2}=-a(\rho)^{2}dt^{2}+d\rho^{2}+b(\rho)^{2}d{\bf x}_{D-2}^{2}\label{metrick0}
\end{align}
%where $n=D-2$.
with
%the
$(D-2)$-dimensional planar metric
%coordinates
$d {\bf x}^2_{D-2}$.
Employing the same {ansatz \eqref{Ansatz} for the} scalar and gauge fields' , it is straightforward to show that the action \eqref{action} is invariant under the following finite scaling transformations
\begin{\eq}
a(\rho)&\rightarrow &\lambda^{2-D}a(\rho)\label{tr1},\\
b(\rho)&\rightarrow &\lambda b(\rho)\label{tr2},\\
A_{t}(\rho)&\rightarrow &\lambda^{2-D}A_{t}(\rho)\label{tr3}
\end{\eq}
{with an arbitrary constant parameter $\la>0$. Here, we note that the radial coordinate $\rho$ as well as the scalar field $\vp$ is not transformed and this makes our computation  simpler: If we need a scaling transformation for $\rho$ also, the radial derivatives for the fields do not simply transform as \eqref{tr1}-\eqref{tr3} and this makes the computation more complicated.} Then, to find the Noether charge we use a field theory trick  where
the {global}
%constant
symmetry parameter $\lambda$ is {{\it localized} as}
%upgraded to
%an arbitrary function of $r$, {\it i.e.},
$\la(\rho)$ and then the Noether charge is
%then
the coefficient of $\lambda^{\prime}(\rho)$ in the variation of the action.
The validity of
%extension of
this trick with higher derivatives and ``higher-level" analogs of Noether's
theorem {were}
%was
%can be
%similarly
proved in \cite{Townsend:2016slt} and
%discussed in \cite{Townsend:2016slt} to include higher derivatives,
%where "higher-level" analogs of Noether's theorem can be similarly proved, and
the corresponding Noether charge ${\cal Q}$ {reads}
%off
from
%, e.g.
the coefficient of $\lambda^{\prime\prime}(\rho)$ in the variation of the action.

Computing the transformation of action, we find the conserved scaling Noether
charge
%to be
\cite{Liu:2015}
\begin{\eq}
{\cal Q}&=&\frac{2}{\kappa} (D-2)   \left(a' b^{D-2}-a b^{D-3} b' \right)-\frac{(D-2)b^{D-2} A_{t} A_{t}'Z\left(|\varphi|^2\right) }{\kappa a}\nonumber\\
&&+4 \beta(|\vp|^{2})  (D-4) (D-3) (D-2) b^{D-5} \left(a b'^3-a'b  b'^2\right) \no \\
&&+16 \dot{\be}(|\vp|^{2}) |\vp \vp'| b^{D-4} b' (ab'-a'b).\label{chargek0}
\end{\eq}
The radial derivative of \eqref{chargek0} is combination of field equations which indicates that it is conserved ``on-shell"
\begin{\eq}
{\cal Q}^{\prime}&=&(D-2) \left(a E_{a}  -\frac{b E_{b} }{(D-2)}+ A_{t} E_{A_{t}}  \right),\label{consv1}
\end{\eq}
where $E_{a}, E_{b}, E_{A_{t}}$ are the field equations of $a, b, A_{t}$, respectively.

Here, it is important to note that {the scaling symmetry and its associated
{\it local} charge \eqref{chargek0} is valid only for the planar topology
\cite{Gubs:2009}.} %For example,
{In other words,} if
%consists of only {\it local} terms in accordance with the Noether's theorem
%as the action is invariant under \eqref{tr1}-\eqref{tr3}.
we move onto the spherical (or hyperbolic) topology
%geometry
with the corresponding metric
\begin{\eq}
ds^{2}=-a(\rho)^{2}dt^{2}+d\rho^{2}+b(\rho)^{2}d\Omega_{D-2}\label{metrick1}
\end{\eq}
and
%the
$(D-2)$-dimensional spherical (or hyperbolic) metric
%coordinates
$d\Omega_{D-2}$, one finds that
%Because of the spherical symmetry, this time
the action \eqref{action} is {\it not} invariant under the finite transformations \eqref{tr1}-\eqref{tr3} by the $k$-dependent terms. However, one can still
{\it construct}
%write down
a conserved charge
%quantity
by simply {adding}
%{\it subtracting}
the {{\it space integral}} of all the non-conservation
%conservation-violating
%non-invariant
terms in \eqref{consv1}
%under an integral
{with a minus sign}
\cite{Cai:2020,Deve:2021} so that all the non-conservation terms are canceled. {But in this case, the conserved charge has {\it non-local} ({\it i.e.} integral) terms as well as the usual local terms, and so it seems that there is no relation to the action invariance and it is beyond the Noether's theorem.} Following this idea and a coordinate transformation to our main metric \eqref{Ansatz}, we find the
conserved charge formula
%that is given in
\eqref{Q} with the radial conservation equation (\ref{Q_prime}).

\section{Proof of $A_t=0$ at the horizons for charged GB black holes with charged {(complex)} scalar hairs}
{\label{app3}}

In this Appendix, we present a proof of the condition (\ref{BC_A}), $A_t=0$ at the horizons $r_H$ for charged GB black holes with {\it charged} ({\it i.e.}, complex) scalar hairs. This is one of
the key ingredients in {our}
%the
``no scalar-haired Cauchy horizon theorem".

To this end, we first note that `$q \varphi A_t|_{r_H}=0$' from the regularity
({\it i.e.}, no singularities) of the equations of motion (\ref{E_A})
%,\ref{E_varphi},\ref{E_N},
-~(\ref{E_f}) at the horizons $r_{H}$. [Here, we consider $D=4$ case for
simplicity but the result is unchanged for arbitrary higher dimensions also]
Then, in order to avoid the horizon singularities in (\ref{E_A})
%,\ref{E_varphi},\ref{E_N},
-~(\ref{E_f}) for the {\it charged} scalar field ($q \neq 0$), we need to consider either (a) $A_t|_{r_H}=0$ or (b) $\varphi|_{r_H}=0$ with $A_t|_{r_H} \neq 0$. We will prove that the case (a) is the only possible condition for {\it hairy} black holes with a {\it smooth} scalar field $\vp$.

%To that end, we take
{Aiming for} the proof by contradiction \cite{Cai:2020},
%and so
let us first suppose that the case (b) is true. From the smoothness of $\varphi$, one can  consider the Taylor expansion near the horizons $r_H$,
\begin{\eq}
\varphi(r)=\varphi_m \de^m + \varphi_{m+1} \de^{m+1} +\cdots
\end{\eq}
with $\varphi_m \equiv \f{1}{m !} \varphi^{(m)}|_{r_H} \neq 0$ ($m \geq 1$) and $\de\equiv r-r_H$. Similarly, from the smoothness of $\sqrt{g} \sim \sqrt{f/h}$ at the horizons, one can consider the metric functions' expansions,
\begin{\eq}
h(r)&=&h_{n}\de^n + h_{n+1} \de^{n+1} +\cdots, \\
f(r)&=&f_{n}\de^n +f_{n+1} \de^{n+1} +\cdots
\end{\eq}
with $h_{n}, f_{n} \neq 0$ ($n \geq 1$). From the smoothness of the scalar equation (\ref{E_varphi}), we need $m \geq n$ and then (\ref{E_varphi}) becomes
\begin{\eq}
2 \alpha r^2_{H}  \varphi_m  \left( q^2 A_t^2|_{r_H}+m^2 f_1  h_{1} \right)  +{\cal O}(\de)=0, \label{Taylor_Eq}
\end{\eq}
by considering {the}
%a
non-extremal black hole, {\it i.e.}, $n=1$, for simplicity.
Then, it is easy to see that there is no way to satisfy (\ref{Taylor_Eq}) for the case (b) with $\vp_m \neq 0$ at the leading order and for the Killing horizons with $f_1 \sim h_1$, up to a {\it positive} factor, from the surface gravity $\ka_H \sim f'|_{r_H}\sim h'|_{r_H}$. Actually, the resulting leading terms in (\ref{Taylor_Eq}) are the same as in EMS theories, even though there are GB corrections in the subleading terms. This proves that the {case} (b) can not be correct and the case (a) is the only correct condition \footnote{Recently, we became aware of an earlier discussion on the condition (a) in a different context \cite{Hong:2019}.}. All these proofs are unchanged for any higher value of $m \geq n$ and arbitrary higher dimensions $D \geq 4$, where (\ref{E_varphi_Full}) becomes
\begin{\eq}
 && 2 \alpha r^2_{H}  r^{D-4}_{H}  \varphi_n   \left[ q^2 A_t^2|_{r_H} +m(m+n-1) f_n h_n \de^{2m-2}  \right] +{\cal O}(\de^n)=0. \label{Taylor_Eq_Full}
\end{\eq}

\section{Proof on general behaviors of $A_t(r)$ with $A_t|_{r_H}=0$}
{\label{app4}}

\begin{figure}
\includegraphics[width=8cm,keepaspectratio]{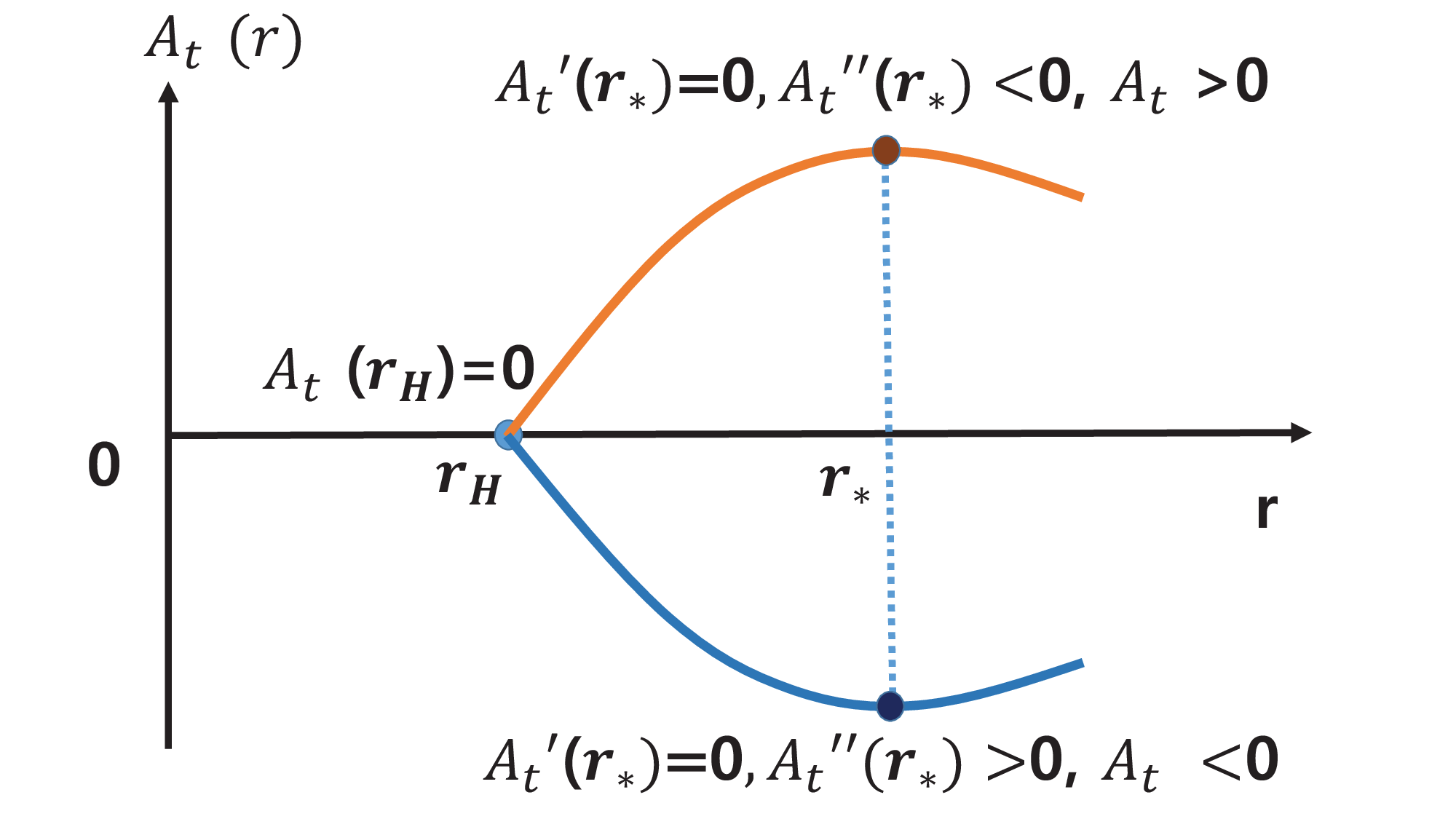}
\includegraphics[width=8cm,keepaspectratio]{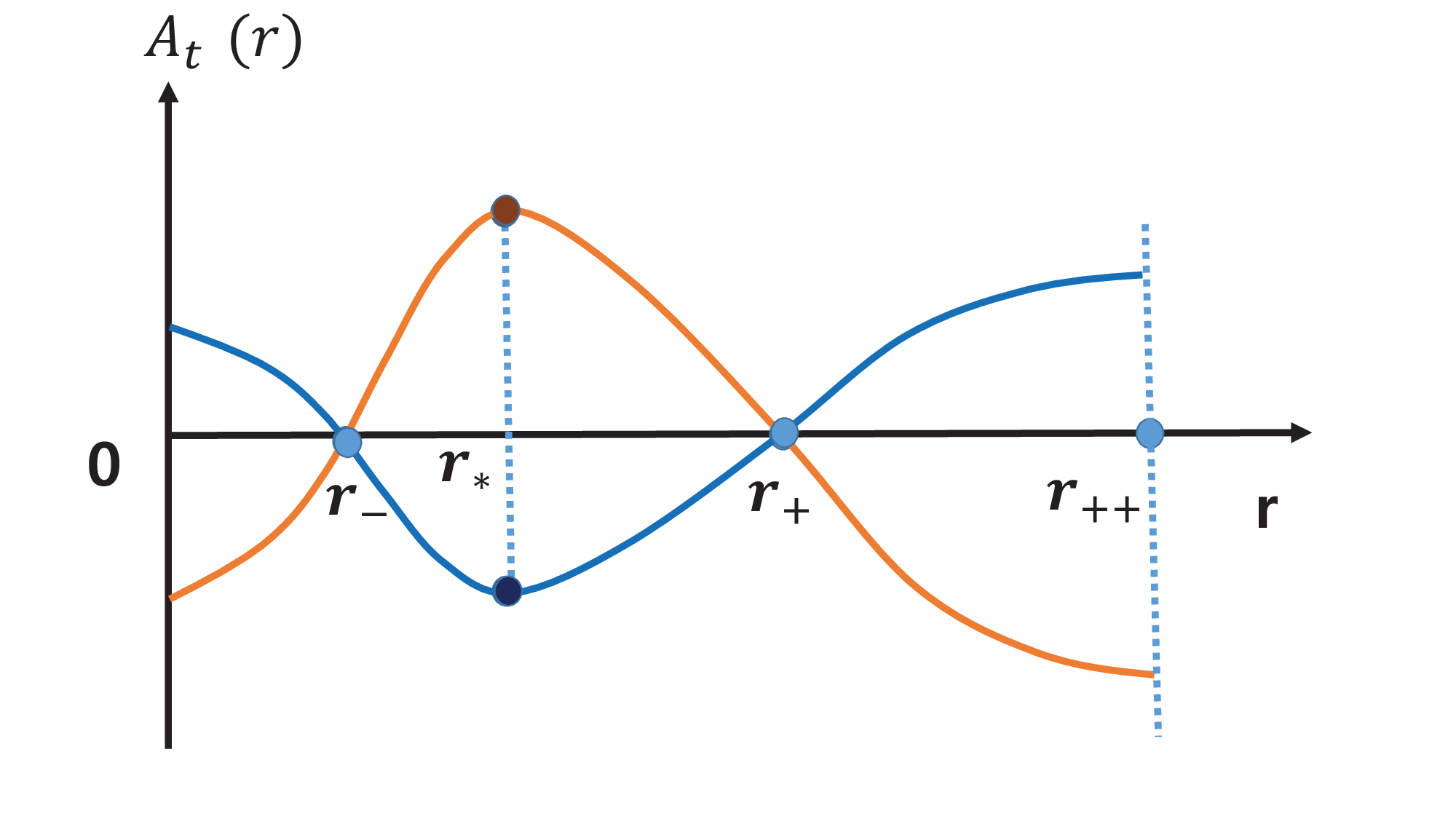}
\caption{{In the left panel, we plot the} behaviors of $A_t(r)$ with extremum points $r_*$ and $A_t|_{r_H}=0$ at the horizons $r_H$. {In the right panel, we consider the generic case with the inner horizon $r_-$ and the cosmological horizon $r_{++}$, as well as the outer horizon $r_+$.} When there is a cosmological horizon in de Sitter black holes, $A_t(r)$ can not be zero due to the monotonicity property in the black hole exterior region $r \in [r_+,r_{++}]$ from (\ref{A_max}). But this is in contradiction to the property of $A_t|_{r_H}=0$ at the horizons $r_H$.
}
\label{fig:At1}
\end{figure}

In this Appendix, we present a proof on general behaviors of $A_t (r)$ when
%independently on
the property of $A_t|_{r_H}=0$ (\ref{BC_A}) {is satisfied} for charged scalar hairs with $q\neq 0$.

We first prove that ``there is no local extremum of $A_t(r)$ for the time-like
region with $f,h >0$". In order to prove this, we first consider a maximum point
$r_*$ where
%with
$A_t'|_{r_*}=0,~A_t''|_{r_*}<0$ are satisfied. Then, the gauge field equation (\ref{E_A}) becomes
\begin{\eq}
A_t''|_{r_*}=\left.\f{2 \alpha q^2 A_t^2 |\varphi|^2 }{f Z}\right|_{r_*}
\label{A_max}
\end{\eq}
at the extremum point $r_*$ in arbitrary dimensions \footnote{(\ref{A_max}) has {\it no} GB effect and it is the same as that of \cite{Cai:2020}. But, with the Einstein coupling (constant) $\ga$ in Horndeski theories, (\ref{A_max}) becomes   $A_t''|_{r_*}=\left.\f{2 q^2 A_t^2 |\varphi|^2 }{r^2 f Z}[\alpha r^2+\ga (k-f)-\ga f f']\right|_{r_*}$ in $D=4$, for simplicity, and we have the same result on the extremum of $A_t$ if $[\alpha r^2+\ga (k-f)-\ga f f']>0$ is satisfied.}. If we
consider a positive $A_t(r)$ in {the}
%a
region starting from $r_H$ (Fig. {\ref{fig:At1} (left), upper curve) with
$A_t|_{r_H}=0$ (\ref{BC_A}) (Appendix {\bf C}), then it is easy to find that
the maximum point $r_*$ with $A_t''|_{r_*}<0 ~( A_t>0)$ is not possible
unless $f,h<0$, {\it i.e.}, {the}
%a
space-like region since we have assumed charged scalar fields with $q \neq 0$ and a non-minimal coupling factor $Z(|\vp|^2)>0.$

Similarly, one can also prove that there is no local minimum point $r_*$ with $A_t'|_{r_*}=0,~A_t''|_{r_*}>0$
for a negative $A_t(r)$ in {the}
%a
region starting from $r_H$ with $A_t|_{r_H}=0$, unless $f,h<0$ (Fig. {\ref{fig:At1} (left), lower curve). Our proof is essentially the same as that of \cite{Cai:2020} but with a simple manner in our context.

So, we have proved that $A_t$ can have {a local} extremum only in the black hole interior region $r \in [r_-,r_+]$ (Fig. {\ref{fig:At1} (right)), in order to be consistent with the condition of $A_t|_{r_H}=0$ (\ref{BC_A}) for charged scalar hairs. We have confirmed this property in the numerical solutions for asymptotically {anti-}de Sitter space in Sec. V.

One remarkable corollary is{, as noted in Sec. III,} that ``there is no {\it de Sitter} (dS) black holes with
charged scalar hairs in EMS or EMGBS theories in arbitrary dimensions" and it
%which
can be simply proved as follows. First, suppose that we apply the above property
on the extremum of $A_t$ to the black hole exterior region $r \in [r_+,r_{++}]$
for the black hole horizon $r_+$ and {the} cosmological horizon $r_{++}$, then
one easily find that $A_t$ needs to be monotonically increasing or decreasing,
depending on its value in the interior region $r \in [r_-,r_{+}]$
(see Fig. {\ref{fig:At1} (right)). Then, $A_t$ at the cosmological horizon
$r_{++}$ will not be vanished but this is in contradiction to the property
of $A_t|_{r_H}=0$ (\ref{BC_A}) for charged scalar hairs. This proves the
above {\it no-go} theorem for charged dS black holes with charged scalar
hairs \footnote{For a more general proof in $D=4$, see \cite{An:2023}.}.
Of course, this does not exclude the neutral{, {\it i.e.,} real} scalar hairs
for charged dS or charged scalar hairs for
%a
{neutral} dS black holes. Another interesting possibility of evasion of the no-go theorem is the case with Einstein coupling in Horndeski theories (see footnote No. 4) if $[\alpha r^2+\ga (k-f)-\ga f f']<0$ is satisfied in the region $r \in [r_+,r_{++}]$.

\section{Relevant computations on
%the regularity relations at the horizons
(\ref{C=0})
%(19)
% and (20)
}
{\label{app5}}
In this Appendix, we
%will
present the computational details on the relation (\ref{C=0})
%and (20)
in Sec. IV.
%tion
%$\ref{section4}$.
Here, for the sake of computational simplicity and practical purpose in the
numerical solutions, we
%will
consider {the $D=4$ case}.
%Since we have  discussed the numerical solutions for the four dimensional case,
%our discussion here will be limited to four dimensions.
Let us
%We
begin by considering the functions $P,S,U,W,R,Y$ given in \eqref{PUWRSY}
\begin{\eq}
P&\equiv& -2 \alpha  r^2 e^{A+4 B} \left(r^2 \kappa \left(2 e^{B}+\mu\right) V-4   \left(k e^{B}-1\right)\right)\no\\
&&+2 r^2 e^{3 B } \Big\{-2 \alpha  q^2 \kappa e^{B} \vp^2 A_{t}^2 \left(\alpha  r^2 e^{B}-16 \left(k e^{B}-1\right) \dot{\beta}\right)\no\\
&&-  A_{t}'^2 \left(\alpha  r^2 e^{B} Z-16 \vp^2 \left(k e^{B}-1\right) \dot{\beta} \dot{Z}\right)\Big\}+\no\\
&&-4 e^{A} \Big\{16 \kappa  \vp^2 \left(k e^{B}-1\right) \Big(A' \left(k e^{B}-3\right) \nu' \dot{\beta}^2+2 e^{2 B} A'^2 \left(k e^{B}-1\right) \dot{\beta}^2+2 \alpha  r^2 e^{3 B} \vp'^2 \ddot\beta+r^2 e^{4 B} \dot\beta \dot V\Big)\no\\
&&-e^{B} \mu ' \left[{\alpha  r^3 e^{2 B}}+8 \kappa  \vp \dot\beta \left(2 \vp A' \left(k e^{B}-3\right) \left(k e^{B}-1\right) \dot\beta-\alpha  r^2 e^{B} \vp'\right)\right]\no\\
&&-8 \alpha  \kappa  r
%e^{B}
\vp \vp' \dot\beta
\left({e^{2B} (k e^B-1)\left(r A'+4\right)}+r \nu'\right)
%\left(k e^{3 B} \left(r A'+4\right)
%-e^{2 B} \left(r A'+4\right)+r \nu'\right)
\no\\
&&+\alpha  r^2 e^{2 B} \left[\kappa e^{B} \vp '^2 \left(16 \left(k e^{B}-1\right) \dot\beta +\alpha  r^2 e^{B}\right)+r \nu'\right]\Big\}\no\\
&&+e^{B} \mu \Big\{2 \kappa  e^{A} \Big[-16 \vp^2 \left(k e^{B}-1\right) \left(2 A'^2 \left(k e^{B}-1\right) \dot\beta^2+2 \alpha  r^2 e^{B} \vp'^2 \ddot\beta+r^2 e^{2 B} \dot\beta \dot V\right)\no\\
&&+8 \alpha  r e^{B} \vp  \left(r A'+4\right) \left(k e^{B}-1\right) \vp'\dot\beta -\alpha  r^2 e^{B} \vp'^2 \left(16 \left(k e^{B}-1\right) \dot\beta+\alpha  r^2 e^{B}\right)\Big]\no\\
&&+4 \alpha  r^2 e^{A+2 B} \left(k e^{B}-1\right)+r^2 e^{B} \Big[-2 \alpha  \kappa  q^2 e^{B} \vp^2 A_{t}^2 \left(\alpha  r^2 e^{B}-16 \left(k e^{B}-1\right) \dot\beta\right)\no\\
&&-A_{t}'^2 \left(\alpha  r^2 e^{B} Z-16 \vp^2 \left(k e^{B}-1\right) \dot\beta \dot Z\right)\Big]\Big\},\\
S&\equiv& 128 \kappa  \vp^2 e^{A+B} \left(k e^{B}-1\right)^2 \left(2 e^{B}+\mu\right) \dot\beta^2, \\
U &\equiv & -2 \kappa  r^2 e^{A+3 B} \left(2 e^{B}+\mu\right) V-r^2 e^{2 B} \left(2 e^{B}+\mu\right) \left(2 \alpha  \kappa  q^2 e^{B} \vp^2 A_{t}^2+Z A_{t}'^2\right)\no\\
&&+2 e^{A} \Big\{e^{B} \mu  \left[\kappa  \vp'^2 \left(-32 \vp^2 \left(k e^{B}-1\right) \ddot\beta-16 \left(k e^{B}-1\right) \dot\beta-\alpha  r^2 e^{B}\right)+2 e^{B} \left(k e^{B}-1\right)\right]\no\\
&&-2 e^{3 B} \left[\kappa  \vp'^2 \left(16 k \left(2 \vp^2 \ddot\beta +\dot\beta \right)+\alpha  r^2\right)+2\right]-2 e^{B} \mu' \left(4 \kappa  \vp \left(k e^{B}-3\right) \vp' \dot\beta+r e^{B}\right)\no\\
&&-2 e^{B} \nu' \left(4 \kappa  k \vp \vp' \dot\beta +r\right)+4 k e^{4 B}+32 \kappa  e^{2 B} \vp '^2 \left(2 \vp^2 \ddot\beta+\dot\beta \right)+24 \kappa  \vp\nu' \vp' \dot\beta \Big\},\\
W&\equiv& 32 \kappa  \vp  e^{A+B} \left(k e^{B}-1\right) \left(2 e^{B}+\mu \right) \dot\beta,\\
R&\equiv & 4 \alpha  \kappa  q^2 r e^{2 B} \vp^2 A_{t} \left(2 e^{B}+\mu \right)
+A_{t}'\Big\{Z \left[e^{B} \left(\mu \left(r A'-4\right)-r \mu'\right)+2 e^{2 B} \left(r A'-4\right)-r \nu'\right]\no\\
&&-4 r e^{B} \vp \left(2 e^{B}+\mu \right) \vp' \dot{Z}\Big\},\\
Y&\equiv& 2 r e^{B} \left(2 e^{B}+\mu\right) Z,
\end{\eq}
where $e^B,\mu,\nu$ are functions of $A,\vp,A_t$ and their first radial
derivatives as given in (\ref{eB})-(\ref{nu}), and (\ref{eB'}).
%(13)-(15), (17).
{Without the gauge field $A_t$}, {{\it i.e.}, for {\it neutral} GB black holes
$(\widetilde{Q}=0)$,} {the above results correspond to those of
%Antoniou et. al.
\cite{Anto:2017} for a real scalar field $\phi$ and the coupling parameter $\be=f(\phi)$.}

By expanding {all the above functions} and equations in \eqref{PUWRSY} near the horizons, {\it i.e.} taking $A\ra \infty, A'\ra \pm \infty$ for $r_{\pm}$, we find the aforementioned constraint (\ref{C=0}),
$C(\vp,\vp^\prime,A_{t}^{\prime},T) \equiv C_{1}/C_{2}=0$
with
%where $C_{1}$ and $C_{2}$ read
\begin{\eq}
C_{1}&\equiv&\left(4 \kappa  k \vp \vp' \dot\beta +r\right)\Big\{ 256 \kappa ^3 T^2 \vp^3 \vp'^2 \dot\beta^2 \left(r^2 \dot V-4 \kappa  k V \dot{\beta} \right)\no\\
&&+16 \kappa  k T \vp^2 \vp' \dot\beta  \Big[8 \kappa  r T \left(\kappa  V \dot\beta  \left(\kappa  r^2 V-6 k\right)+r^2 \dot{V}\right)\no\\
&&+A_{t}'^2 \left(r^2 \dot{Z} \left(\kappa  r^2 V-2 k\right)+2 \kappa  Z \left(2 \dot\beta  \left(2-3 \kappa  k r^2 V\right)+r^4 \dot{V}\right)\right)\Big]\no\\
&&+2 \alpha  \kappa  r^3 T \vp' \left(\kappa  r^2 V-2 k\right) \left(r Z A_{t}'^2+4 T\right)\no\\
&&+\vp\Big(-8 \kappa  \dot{\beta}\Big[r^2 \Big(\kappa  k V \left(-4 T^2 \left(\alpha  \kappa  r^2 \vp'^2+2\right)+r^2 Z^2 A_{t}'^4+4 r T Z A_{t}'^2\right)+2 \kappa ^2 r^2 T^2 V^2\no\\
&&+8 \alpha  \kappa  T^2 {\vp'}^2-Z^2 A_{t}'^4\Big)+24 T^2\Big]+r^3 A_{t}'^2 \dot Z \left(\kappa  r^2 V-2 k\right) \left(r Z A_{t}'^2+4 T\right)+\kappa  r^4 \dot V \left(r Z A_{t} '^2+4 T\right)^2\Big) \Big\}\no,\\
C_{2}&\equiv &2 \kappa  T \Big\{-2 \alpha  k r^4 \left(r Z A_{t}'^2+4 T\right)+128 \kappa ^3 r^3 T \vp^2 V^2 \dot\beta^2 \left(k r-2 \kappa  \vp  \vp' \dot\beta \right)\no\\
&&+\kappa  r V \left(16 \kappa  k T \vp \vp' \dot\beta  \left(\alpha  r^4+32 \kappa  \vp^2 \dot\beta^2\right)+\alpha  r^5 \left(r Z A_{t}'^2+4 T\right)-32 \kappa  r \vp^2 \dot\beta^2 \left(r Z A_{t} '^2+20 T\right)\right)\no\\
&&+64 \kappa  k \vp^2 \dot\beta^2 \left(r Z A_{t}'^2+12 T\right)-32 \alpha  \kappa  r^3 T \vp \vp' \dot\beta \Big\}
\end{\eq}
{from the finite $\vp''$ at the horizons.} {Here, we note that the results on $C_1$ and $C_2$ depends on the value of the topology parameter $k$ and where we evaluate, as $e^B$ does in (\ref{eB}). In the above, we showed the negative root case for {the $k=-1$ case} and the outer horizon of the black hole interior.}
Now, by solving the constraint $C_1=0$ ($C_2\neq 0$ generically) in terms of $\vp'$, we find three roots
\begin{\eq}
\vp'_{\pm}&\equiv &D_{\pm}/{\tilde D},\\
D_{\pm}&\equiv& 4 \kappa  r T^2 \left[\kappa  \left(16 k r^2 \vp^2 \dot\beta  \dot V+16 \kappa  k \vp^2 V \dot\beta^2 \left(\kappa  r^2 V-6 k\right)+\alpha  r^4 V \right)-2 \alpha  k r^2\right]\no\\
&&+\kappa  T A_{t}'^2 \Big[8 k r^2 \vp^2 \dot\beta  \left(\dot Z \left(\kappa  r^2 V-2 k\right)+2 \kappa  r^2 Z \dot V \right)+32 \kappa  k \vp^2 Z \dot\beta^2 \left(2-3 \kappa  k r^2 V \right)\no\\
&&+\alpha  r^4 Z \left(\kappa  r^2 V -2 k\right)\Big]\pm\Bigg\{\kappa ^2 T^2 \Big[\Big(-8 \alpha  k r^3 T+4 \kappa  r T \Big( 16 k r^2 \vp^2 \dot\beta \dot V+16 \kappa  k \vp^2 V \dot\beta^2 \left(\kappa  r^2 V-6 k\right)\no\\
&&+\alpha  r^4 V \Big)+A_{t}'^2 \Big[8 k r^2 \vp^2 \dot\beta \left(\dot Z\left(\kappa  r^2 V-2 k\right)+2 \kappa  r^2 Z \dot V\right)+32 \kappa  k \vp^2 Z \dot\beta^2 \left(2-3 \kappa  k r^2 V\right)\no\\
&&+\alpha  r^4 Z \left(\kappa  r^2 V-2 k\right)\Big]\Big)^2-32 \vp^2 \dot\beta  \Big(\kappa  k V \left(\alpha  r^4-32 \kappa  \vp^2 \dot\beta^2\right)-2 r^2 \left(\alpha -4 \kappa  \vp^2 \dot\beta  \dot V \right)\Big)\no\\
&&{\times}\Big[-8 \kappa  \dot\beta  \left[r^2 \left(2 \kappa  T^2 V \left(\kappa  r^2 V-4 k\right)+Z^2 A_{t}'^4 \left(\kappa  k r^2 V-1\right)+4 \kappa  k r T V Z A_{t} '^2\right)+24 T^2\right]\no\\
&&+r^3 A_{t} '^2 \dot Z \left(\kappa  r^2 V-2 k\right) \left(r Z A_{t}'^2+4 T\right)+\kappa  r^4 \dot V \left(r Z A_{t}'^2+4 T\right)^2\Big]\Big]\Bigg\}^{1/2},\\
\tilde{D}&\equiv& 32 \kappa ^2 T^2 \vp \dot\beta  \left(\kappa  k V \left(32 \kappa  \vp^2 \dot\beta^2-\alpha  r^4\right)+2 r^2 \left(\alpha -4 \kappa  \vp^2 \dot\beta \dot V\right)\right),\\
\vp'_{3}&=&-\frac{r}{4 \kappa  k \vp \dot\beta}.
\end{\eq}
{Here, we have considered a rapidly vanishing gauge field $A_t$ at the horizons so that $e^{-A} |\vp|^2 A_t^2$ term can be neglected to solve the condition $C_1=0$, which is quite involved when we keep the mentioned term.}
Note that these solutions are valid only
%found
at the horizons $r_H$, {\it i.e.,} the outer event horizon $r_+$
or the inner Cauchy horizon $r_-$ and so they can be used as the initial
conditions for the numerical solutions.
%therefore $r=r_{H}$.
In our numerical solutions, we have considered only the {\it positive} root
$\vp'_{+}$ because it only shows
%In this case, only the positive root $\vp'_{1}$ is allowed since it exhibits
a smooth Einstein limit as $\beta\rightarrow 0$.
%%%%%%%%%% References %%%%%%%%%%%%%%%%%%%%%%%%%
\newcommand{\J}[4]{#1 {\bf #2} #3 (#4)}
\newcommand{\andJ}[3]{{\bf #1} (#2) #3}
\newcommand{\AP}{Ann. Phys. (N.Y.)}
\newcommand{\MPL}{Mod. Phys. Lett.}
\newcommand{\NP}{Nucl. Phys.}
\newcommand{\PL}{Phys. Lett.}
\newcommand{\PR}{Phys. Rev. D}
\newcommand{\PRL}{Phys. Rev. Lett.}
\newcommand{\PTP}{Prog. Theor. Phys.}
\newcommand{\hep}[1]{ hep-th/{#1}}
\newcommand{\hepp}[1]{ hep-ph/{#1}}
\newcommand{\hepg}[1]{ gr-qc/{#1}}
\newcommand{\bi}{ \bibitem}
%%%%%%%%%%%%%%%%%%%%%%%%%%%%%%%%%%%%%%%%%%%%%%%

\end{document}